\documentclass[twocolumn]{aastex63}

\renewcommand{\edit}[1]{\textcolor{red}{#1}}

\def\nustar{{\it NuSTAR\/}}

\def\xmm{{\it XMM-Newton\/}}
\def\chandra{{\it Chandra\/}}
\def\flux{erg~s$^{-1}$~cm$^{-2}$}

\def\Suzaku{{\it Suzaku\/}}

\def\VLA{{\it VLA}}
\newcommand{\mach}{$\mathcal{M}$}
\usepackage{lineno}
\graphicspath{{./}{figures/}}
\usepackage{amsmath}

\shorttitle{Unlocking Pandora's Secrets: Probing the Thermodynamics of Abell 2744}
\shortauthors{Rojas et al.}

\begin{document}

\title{Unlocking Pandora's Secrets: Probing the Thermodynamics of Abell 2744}

\author [0000-0002-8882-6426]{Randall A. Rojas Bolivar}
\affil{Department of Physics \& Astronomy,
University of Utah,
115 South 1400 East, Salt Lake City, UT 84112, USA}
\affiliation{Center for Astrophysics, Harvard \& Smithsonian, Cambridge, MA 02138, USA}

\author[0000-0003-0573-7733]{\'Akos Bogd\'an}
\affiliation{Center for Astrophysics, Harvard \& Smithsonian, Cambridge, MA 02138, USA}

\author[0000-0001-9110-2245]{Daniel R. Wik}
\affil{Department of Physics \& Astronomy,
University of Utah,
115 South 1400 East, Salt Lake City, UT 84112, USA}

\author[0000-0002-4962-0740]{Gerrit Schellenberger}
\affiliation{Center for Astrophysics, Harvard \& Smithsonian, Cambridge, MA 02138, USA}

\begin{abstract}

Abell 2744 (Pandora's Cluster) is a massive galaxy cluster undergoing multiple mergers, characterized by a highly disturbed X-ray morphology, elevated intracluster medium (ICM) temperatures, and several radio relics associated with merger driven shocks. We present \nustar\ observations of Abell 2744 aimed at mapping its thermodynamic structure and constraining the presence of inverse Compton (IC) scattering from relativistic electrons. We measure the temperature distribution across the cluster and identify a prominent shock front with a Mach number of $\mathcal{M} = 1.58^{+0.42}_{-0.30}$, aligning with the southeastern boundary of the radio halo. The global NuSTAR spectra are modeled with single- and two-temperature thermal components, as well as thermal plus power-law models to test for IC emission. The spectra are well described by purely thermal models with an average temperature of $kT \approx 11$ keV, and we find no significant evidence for non-thermal hard X-ray emission. We derive a 90\% confidence upper limit on the $20-80$ keV IC flux of $ F_{\rm nt} < 2.31 \times 10^{-12}$~\flux. Combining this with 1.5~GHz diffuse radio measurements from the \VLA, we estimate a lower limit on the average magnetic field strength of $B > 0.2 \ \rm{\mu G}$.

\end{abstract}

\keywords{galaxies: clusters: general --- galaxies: clusters: individual (Abell 2744) --- intergalactic medium --- magnetic fields --- radiation: non-thermal --- X-rays: galaxies: clusters}

\section{Introduction}
\label{sec:intro}

Galaxy clusters are the most massive gravitationally bound systems in the known Universe, growing hierarchically through mergers of smaller groups and clusters. Merger events drive shocks, turbulence, and bulk motions within the intracluster medium (ICM). These physical processes heat the plasma, (re)accelerate preexisting fossil electrons via diffusive shock acceleration, and amplify magnetic fields across the galaxy cluster \citep{reacceleration, VanWeeren2019}.

Plasma heating occurs as shock fronts compress the ICM, raising the local gas temperature and creating a temperature jump across the shock. Shock fronts are therefore visible in the X-ray band as both surface brightness and temperature discontinuities. X-ray observations of shock fronts are therefore a useful way to calculate the velocity of the shock-propagated gas \citep{1999ApJ...521..526M}. The Mach number (\mach) can be derived from the temperature jump applying the Rankine-Hugoniot jump conditions. Typical values of \mach\ in merging clusters are between 1 and 3 \citep{Sarazin}. In addition to heating the ICM, these weak shocks play an important role in driving non-thermal emission within the galaxy cluster, particularly through the acceleration of relativistic particles.

Given their low Mach numbers, these shock fronts are expected to be inefficient accelerators, making directly accelerating electrons from the thermal pool difficult \citep{2017JKAS...50...93K}. Many merging clusters exhibit diffuse radio emission in their outskirts, known as radio relics, which indicate the presence of relativistic electrons accelerated by merger shocks. To rectify this apparent contradiction, a population of mildly relativistic fossil electrons must be present to be re-accelerated via diffusive shock acceleration (DSA). These electrons may originate from past active galactic nuclei (AGN) activity or previous shock events, providing a pre-existing population that can be re-accelerated more efficiently than electrons drawn directly from the thermal pool \citep{KangShock,VanWeeren2019}. As they spiral along the magnetic field lines of the galaxy cluster, they emit synchrotron radiation, forming radio relics that can be used to trace the location and morphology of the shock fronts. This synchrotron radiation depends on both the energy density of the electrons and the magnetic field strength. The radio spectral index can also be used to infer the shock Mach number under the assumption of DSA. However, Mach numbers inferred from radio observations can exceed those derived from X-ray temperature or density jumps, highlighting discrepancies between radio-based and X-ray-based estimates of shock properties \citep{Wittor2021}. 

The strength of the magnetic field is determined by a combination of shock compression and turbulent amplification. Compressed gas amplifies existing magnetic fields via flux freezing \citep{Hu}, while turbulence can also enhance magnetic fields with dynamo processes \citep{Roh}. The magnetic fields responsible for synchrotron emission in radio relics also contribute to the emission observed in radio halos, which are diffuse synchrotron sources extending over the central regions of galaxy clusters. Radio halos are generally associated with merger-driven turbulence, which is thought to re-accelerate preexisting relativistic electrons and sustain synchrotron emission over large spatial scales \citep{reacceleration}. Sharp X-ray surface-brightness edges have also been observed near the peripheries of some radio halos, providing additional evidence of structure in the thermal ICM associated with cluster merger activity \citep{Botteon2023}. The magnetic field strength regulates the partition of energy between radio and X-ray non-thermal components within the galaxy cluster. The electrons responsible for radio synchrotron emission should also produce inverse Compton (IC) emission by upscattering cosmic microwave background (CMB) photons into the hard X-ray band. The ratio of synchrotron to IC flux is the only direct observational method to measure the volume-averaged magnetic field strength in galaxy clusters. Accurate measurement of the IC flux is therefore important, as one can derive the non-thermal pressure contribution from the magnetic fields within the cluster to hydrostatic equilibrium, which may explain the biased mass estimates when comparing X-ray to gravitational lensing mass measurements \citep{1993ApJ...407L..49B, 2009astro2010S.305V, 2019A&A...621A..39E}. Direct IC scattering detections have so far proven elusive, with most galaxy clusters yielding upper limits (and therefore lower limits for the magnetic field strength) \citep{rojas21, Rojas665}.

Abell 2744 (A2744), also known as "Pandora's Cluster" due to its many substructures and complex features, is a massive merging cluster ($M_{200} \sim 2 \times10^{15} M_{\odot}$, \citep{2016ApJ...817...24M}) located at a redshift of $z=0.306$ \citep{A2744Owers}. Found at the intersection of several galaxy filaments \citep{2015Natur.528..105E}, A2744 is undergoing a complex four subcluster merger \citep{2011MNRAS.417..333M, A2744Akos}. Radio observations of the cluster show an extended, luminous central radio halo \citep{GovoniA2744} as well as several radio relics up to 2 Mpc away from the center of the cluster \citep{A2744VLA, Paul, RajpurohitA2744}. Some of these radio relics correspond with the presence of shock fronts \citep{MagA2744, A2744VLA, RajpurohitA2744}.

Early \chandra\ observations revealed the complex thermodynamic structure of A2744, with \citet{A2744Owers} and \citet{2011MNRAS.417..333M} reporting a shock front southeast of the cluster core. Subsequent X-ray observations with \xmm\ revealed further evidence of the cluster's complex merger dynamics. In particular, \citet{Jauzac2016} identified four remnant cores and reported three shock features, including a putative shock. Using both \xmm\ and \Suzaku\ observations, \citet{Eckert2016} detected a surface-brightness discontinuity approximately 1.5 Mpc east of the cluster core, coincident with a known radio relic. They inferred a weak shock with a Mach number of $\mathcal{M}=1.7^{+0.5}_{-0.3}$ and measured a post-shock temperature of approximately 13 keV. More recently, deep \chandra\ observations analyzed by \citet{A2744Akos} provided further constraints on the cluster's merger history and shock features.

Radio observations by \citet{A2744VLA} reveal four prominent relics in A2744. The northeastern relic, located approximately 1.5 Mpc from the cluster center, has a clear X-ray counterpart, while a southeastern relic, approximately 1 Mpc from the cluster center, has a weaker X-ray counterpart. The two northwestern relics have no reported X-ray counterparts in that study. \citet{A2744VLA} also discuss the southeastern boundary of the radio halo, which \citet{A2744VLA} identify as aligning with the shock previously reported by \citet{A2744Owers}. This shock is distinct from the weaker X-ray feature associated with the southeastern relic. The global temperature measured with \chandra\ is $kT = 8.77 \pm 0.7$ keV in the 0.1--2.4 keV band \citep{A2744Akos}.

In this paper, we present two deep NuSTAR (Nuclear Spectroscopic Telescope Array; Harrison et al. 2013) observations of A2744 to characterize the thermal properties of the cluster and measure the Mach number across the southeastern shock front originally reported by \citet{A2744Owers}. A second, weaker shock has been identified in the same general region by subsequent X-ray studies \citep{A2744VLA,A2744Akos}, but it lies outside the region where our NuSTAR temperature map provides reliable measurements and is therefore not analyzed in this work. We also constrain the non-thermal inverse Compton (IC) flux using \nustar\ observations, providing the first such measurement for A2744. We combine the resulting IC flux limit with published radio flux measurements at 1.4 GHz from the Very Large Array (VLA) and 240 MHz from the Giant Metrewave Radio Telescope (GMRT) to derive a lower limit on the magnetic field strength \citep{A2744VLA, Paul} In Section~\ref{sec:observe}, we discuss our data reduction procedures. In Section~\ref{sec:analysis}, we discuss how we analyzed the data and determine which model best describes the emission from the galaxy clusters. In Section~\ref{sec:tmap}, we describe our shock front analysis procedure. In Section~\ref{sec:Summary}, we discuss our results. Appendix~\ref{app:BGD} explains in detail background modeling for A2744. 
    
For this paper, all errors shown are at the 90\% confidence level. We assume a flat cosmology with $\Omega_{\rm M}$ = 0.3 and $H_{0} = 70$~km~s$^{-1}$~Mpc$^{-1}$. At the cluster's redshift, 1" is equivalent to 4.5 kpc in the physical scale.

\renewcommand{\footnotesize}{\scriptsize}

\section{Observations and Data Reduction} \label{sec:observe}

\begin{figure*}
\centering
\includegraphics[scale=0.5]{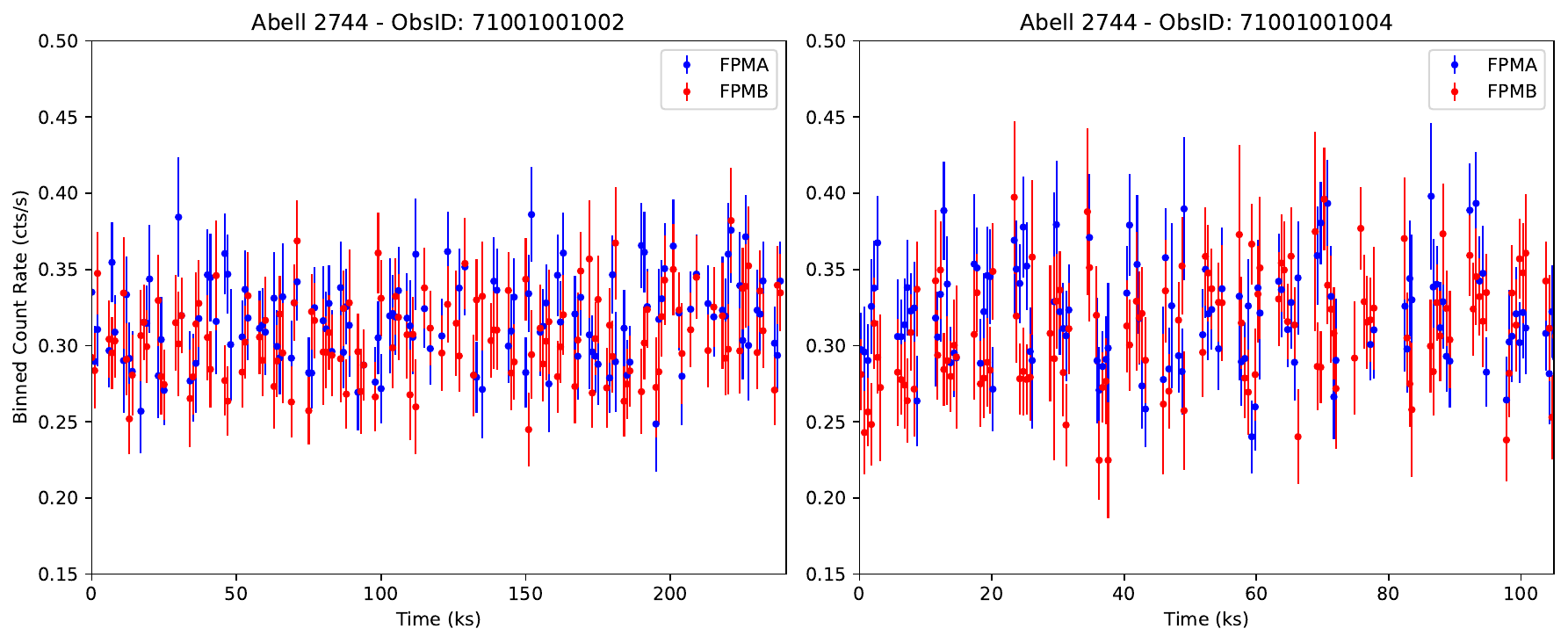}
\caption{
Filtered light curves for the two observations (left: OBSID 71001001002; right: OBSID 71001001004) between 4 and 25 keV following the process described in Section~\ref{sec:observe}. The light curves have been filtered in the $50-160$~keV energy range to eliminate SAA background contributions as well as the $1.6-20$~keV range to remove solar activity background contributions. 
\label{fig:lightcurve}
}
\end{figure*}

 A2744 was observed by \nustar~in two segmented observations between December 21st and December 30th, 2024, the first for a total raw exposure time of 130~ks and the second for a total raw exposure time of 59~ks. These exposures include intervals when the source was occulted by the Earth. We processed the data using the standard {\tt HEASoft} version~6.34 and {\tt NuSTARDAS} version~2.1.4 pipelines. While \nustar's default filtering, which is enabled via the {\tt STRICT} mode and {\tt TENTACLE} flag in {\tt nupipeline}\footnote{https://heasarc.gsfc.nasa.gov/docs/nustar/analysis/nustar\_swguide.pdf}, automatically removes times affected by passages through the South Atlantic Anomaly (SAA), it can be overly strict and remove good data. We therefore removed high background periods from the data manually.

Our manual filtering was achieved by turning off the aforementioned flags and extracting good time intervals (GTIs) using light curves created for both the FPMA and FPMB telescopes by {\tt lcfilter}\footnote{https://github.com/danielrwik/reduc}, with 100~s binning. We visually inspected the light curves and manually excluded the intervals showing elevated count rates when compared to the baseline count rate. This was done in three stages: first in the $50-160$~keV band to exclude high-background periods associated with SAA contamination, then a second pass through in the same energy band to remove residual elevated background, and then a final pass in the $1.6-20$~keV band to account for potential solar activity. After filtering, our exposure time was reduced to 127~ks in the first observation and 54~ks in the second. Figure~\ref{fig:lightcurve} shows the cleaned light curves.

\begin{figure*}
\centering
\includegraphics[scale=0.45]{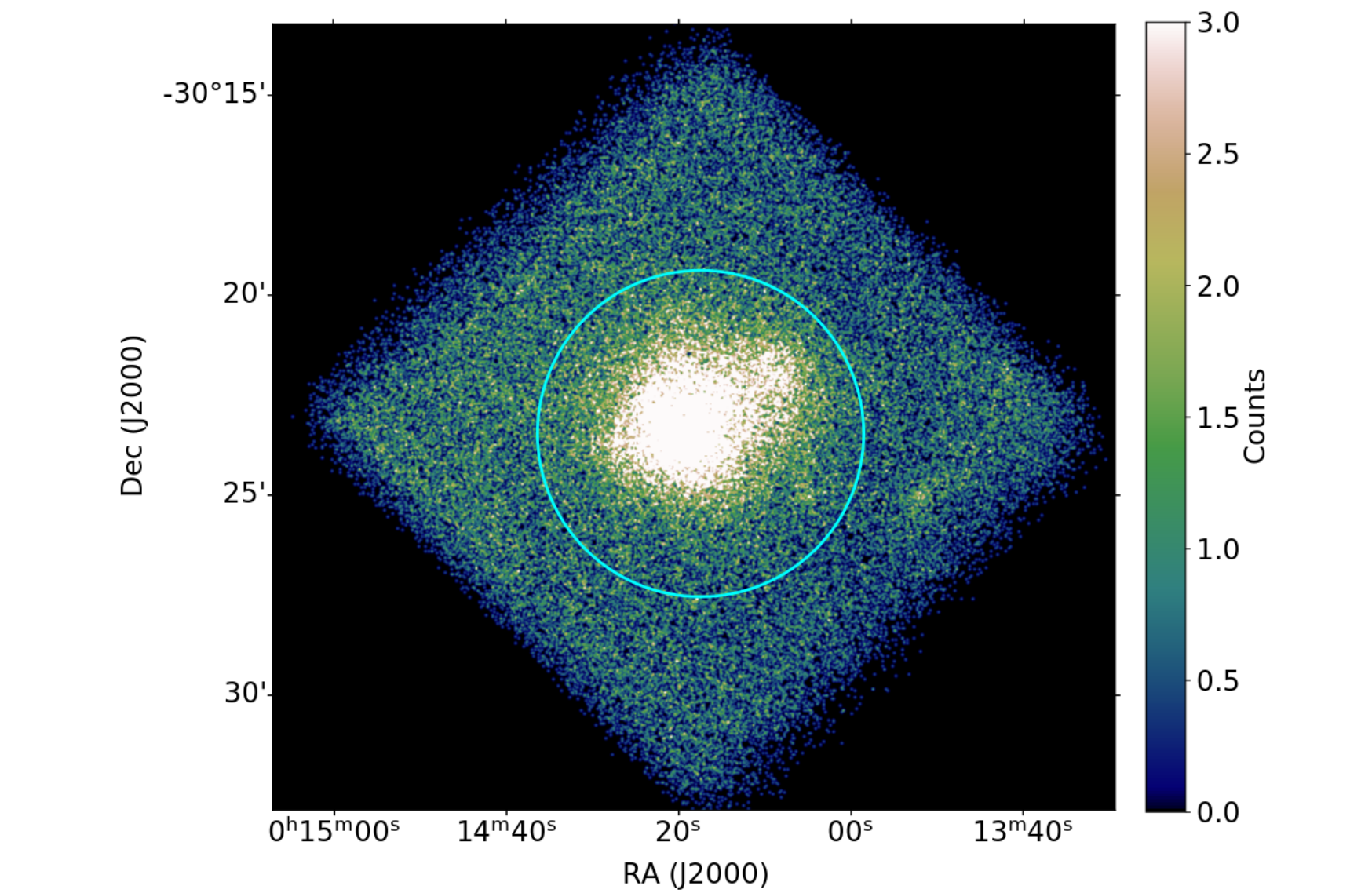}

\caption{A false color (faint to bright represented
by black to blue to pink to white) combined
(A+B for both OBSIDs) log scaled images from $4-25$~keV, smoothed by
a Gaussian kernel with $\sigma$ = 3 pixels, and stretched to show
features in the outer parts of the FOV. The source region from which spectra were extracted is shown as the cyan circle.}
\label{fig:region}
\end{figure*}

We then reprocessed the data using the updated GTIs and generated science products using the standard tools in {\tt nupipeline}.\ Cluster images were created using {\tt XSELECT} post-processing, and exposure maps were generated with {\tt nuexpomap}.\ Spectral products, including source and background spectra, response matrix files (RMFs), and ancillary response files (ARFs), were created with {\tt nuproducts}. The final cleaned and exposure-corrected $4-25$~keV images, along with source region used for subsequent analysis, are shown in Figure~\ref{fig:region}.

Background modelling followed the procedures outlined in \citet{wikBullet} and \citet{rojas21}, with specific details provided in Appendix~\ref{app:BGD}.

\begin{figure*}
\centering
`\includegraphics[scale=0.5]{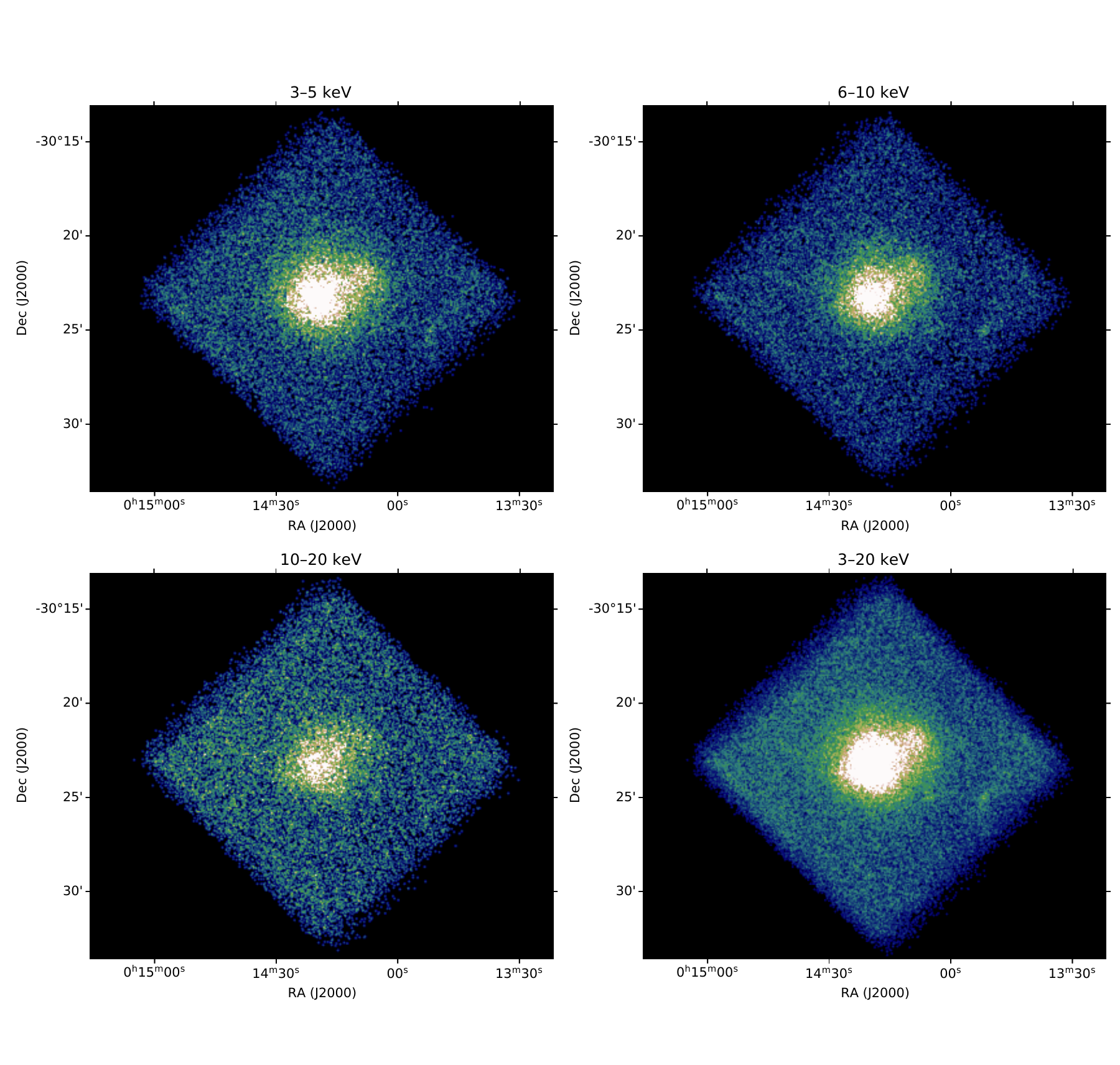}

\vspace{-0.5cm}
\caption{Combined (FPMA+FPMB for both OBSIDs) images in different energy bands. In each panel; top left: 3--5 keV; top right: 6--10 keV; bottom left: $10-20$~keV; bottom right: $3-20$~keV. Each image has been background subtracted and exposure corrected. They are presented in a log scale from 0 counts~s$^{-1}$~pix$^{-1}$ (in black) to 20+ counts~s$^{-1}$~pix$^{-1}$ (in white) and smoothed by a Gaussian kernel with $\sigma$ = 3~pix. 
}
\label{fig:energybands}

\end{figure*}

\section{Global Spectral Analysis} \label{sec:analysis}

To assess whether unresolved point sources could bias our spectral results, we created cluster images in several energy bands by filtering events according to the pulse-height analyzer (PHA) channel using {\tt xselect}. The PHA channel records the measured energy of each detected event, allowing us to examine the spatial distribution of emission over different energy ranges. Figure~\ref{fig:energybands} shows the resulting 3--5, 6--10, 10--20, and 3--20~keV images after exposure correction and background subtraction. The background was modeled with {\tt nuskybgd} (see Appendix~\ref{app:BGD} for details). Examining the images across these energy bands allowed us to check for compact sources that could contaminate the cluster spectrum. We found no prominent point sources likely to bias our spectral measurements.

\subsection{Spectra} \label{subsec:spectra}

We extracted spectra from the region shown in Figure~\ref{fig:region} using {\tt nuproducts}. The $0.8R_{500}$ extraction region was selected to encompass the bulk of the brightest X-ray emission from A2744, providing sufficient signal for spectral fitting while focusing on the main cluster emission. Spectral fitting was performed in {\tt XSPEC}. The full-band spectra for the cluster are presented in Figure~\ref{fig:GlobalSpectra}. The FPMA and FPMB data exhibit excellent agreement, allowing us to carry out joint fits in which all model parameters were tied between the modules except for the normalizations. Spectra were grouped to a minimum of three counts per channel, and we adopted the modified Cash statistic ({\tt statistic cstat}; \citealt{Cash, Cstat}), which retains full Poisson information while avoiding the biases associated with $\chi^{2}$ fitting for low-count X-ray data \citep[e.g.,][]{rojas21}. All spectra were fit between 3 and 20 keV due to low signal-to-noise (S/N) at higher energies.

\subsection{Spectral Models} \label{subsec:models}

Clusters hosting diffuse radio emission, such as A2744, must contain inverse Compton (IC) radiation produced by relativistic electrons scattering CMB photons \citep{Liang}. The challenge is to determine whether this non-thermal emission is strong enough to be detected above the thermal bremsstrahlung spectrum of the intracluster medium (ICM). To assess this, we examined a set of complementary models: purely thermal (1T and 2T) and a hybrid thermal+non-thermal model (T+IC) \citep{2014A&A...562A..60O, wikBullet, rojas21}. All thermal components were described with the {\tt APEC} model (AtomDB v3.0.9), with free temperature, abundance, redshift, and normalization parameters, and using the {\tt wilm} abundance table \citep{Wilm}. Additionally, all models included a cross-calibration constant that hovered around 1 between telescopes. Following \citet{rojas21}, no foreground absorption model was included because \nustar’s effective area rapidly decreases below 3~keV. The redshift was allowed to vary due to \nustar's CALDB file not applying gain shift corrections properly to lines around 6 keV from the nominal 0.308. For a more detailed explanation of why the redshift should be free in the global fit to produce better results, see \citet{Rojas665}.

\subsubsection{Thermal Models} \label{subsubsec:tmodels}

A single-temperature fit provides a reference point for the overall thermal structure, even though a merging cluster such as A2744 is not expected to be isothermal. Merging clusters frequently contain gas that spans a wide range of temperatures, as demonstrated in systems such as Abell~2163 \citep{bourdin2163, rojas21} and Abell~2146 \citep{Helen2146}. Differences in instrument calibration, projection effects, and sensitivity complicate precise measurements of this structure. With \nustar, where line features are limited and the instrument is particularly sensitive to hotter gas, a two-temperature model often offers a better representation of the continuum \citep{rojas21}.
   
\subsubsection{Non-thermal Model} \label{subsubsec:TIC}   

A high-energy non-thermal component may mimic an extremely hot thermal population; therefore, a 2T fit that drives $T_{\rm H}$ to unphysically large values typically motivates consideration of a T+IC model. In this formulation, a single thermal component describes the bulk ICM emission, while a power-law component accounts for putative IC radiation. We adopt a fixed photon index of $\Gamma = 2.43$, derived from the radio spectral index $\alpha \approx 1.43$ \citep{A2744VLA}. Allowing $\Gamma$ to vary introduces degeneracies similar to those encountered with the second temperature component in the 2T model and makes the IC contribution poorly constrained \citep{rojas21}. The temperature and power-law normalization were left free.

\subsection{Fit Results and Systematic Uncertainties} \label{subsec:best}
\subsubsection{Fits to the Global Spectra} \label{subsubsec:globalfitresults}

\begin{figure*}
\centering

\begin{minipage}{1.25\columnwidth}
    \centering
    \includegraphics[width=\linewidth]{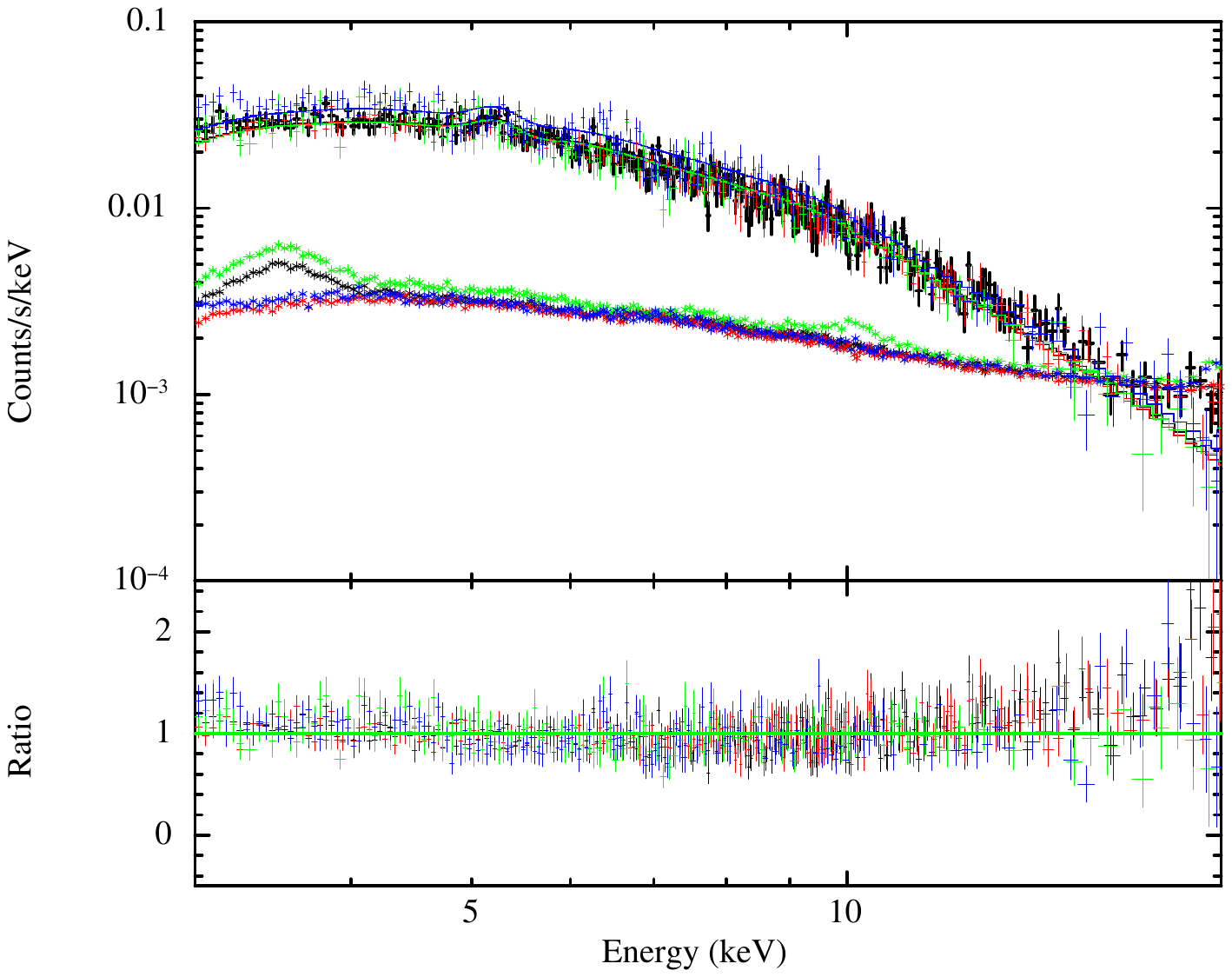}
\end{minipage}
\hfill
\begin{minipage}{1.25\columnwidth}
    \centering
    \includegraphics[width=\linewidth]{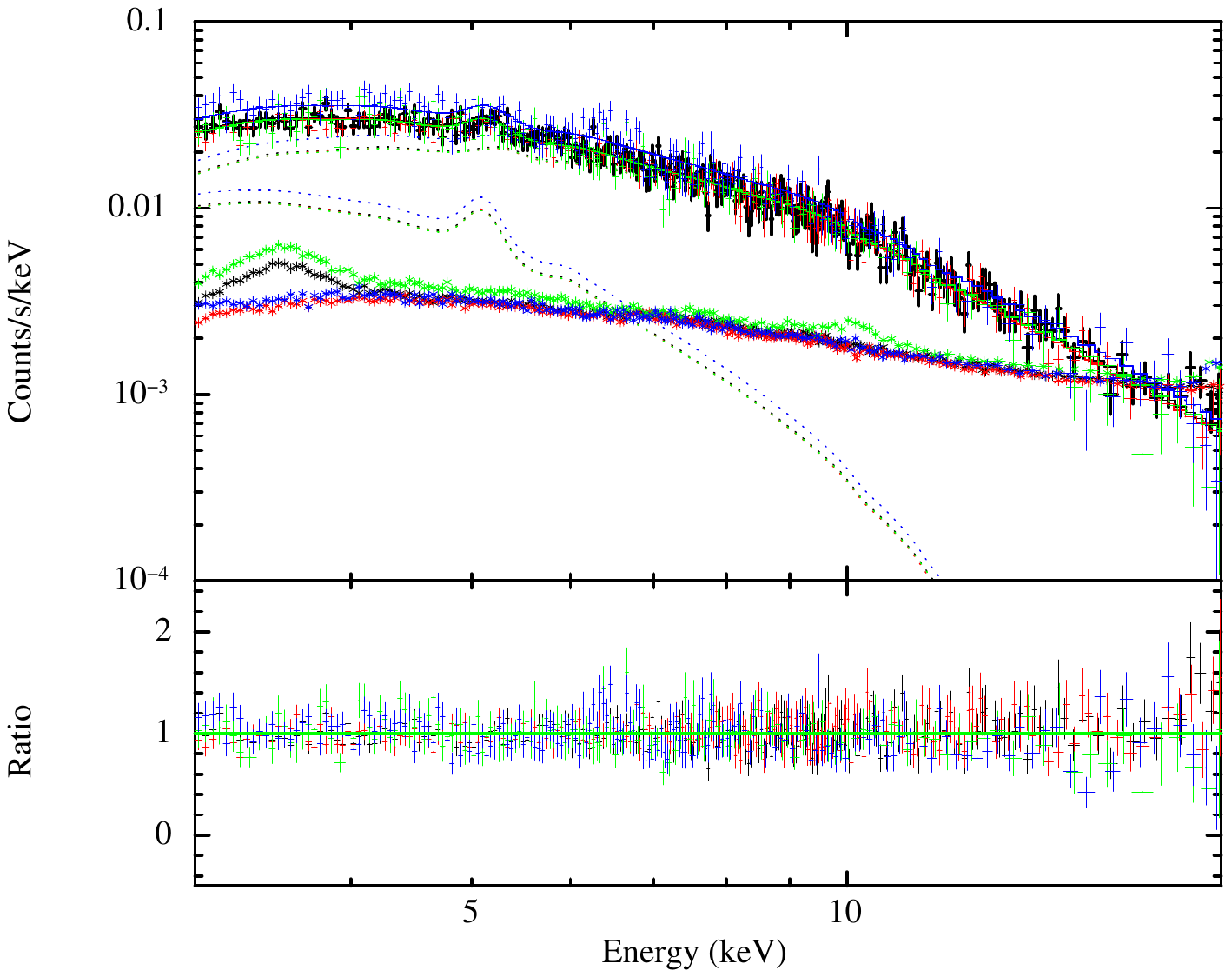}
\end{minipage}
\hfill
\begin{minipage}{1.25\columnwidth}
    \centering
    \includegraphics[width=\linewidth]{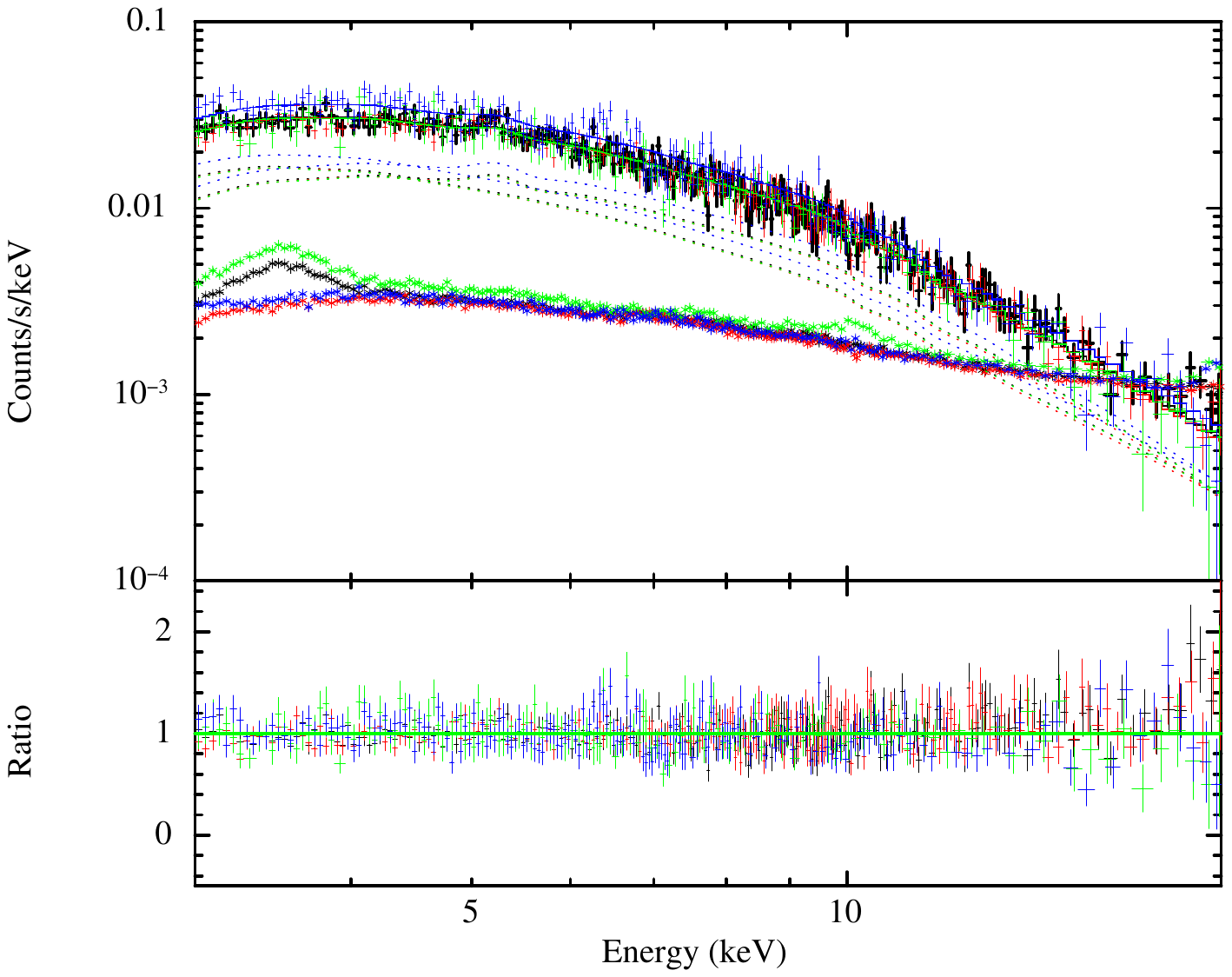}
\end{minipage}

\caption{Global fits to the A2744 spectra with 1T (\textit{top panel}), 2T (\textit{middle panel}), and T+IC (\textit{bottom panel}) models. Black indicates FPMA and red indicates FPMB for OBSID 71001001002, whereas for OBSID 71001001004, FPMA and FPMB are indicated by green and blue, respectively. Crosses show the data and the background is denoted with asterisks. The dashed curves correspond to the model components to visualize their contributions to the composite model. For plotting purposes, adjacent bins are grouped to ensure a detection significance of at least 10$\sigma$, with maximum 20 bins.}
\label{fig:GlobalSpectra}
\end{figure*}

We first compare the results of the 1T and 2T thermal models, followed by the hybrid T+IC model, all of which are shown in Figure~\ref{fig:GlobalSpectra}. The 1T model yields a best-fit temperature of $kT = 10.9 \pm 0.5$~keV (statistical uncertainty only), slightly higher than the \chandra\ measurements of $kT = 9.07 \pm 0.15$~keV \citep{A2744Owers} and $kT = 8.77 \pm 0.7$~keV \citep{A2744Akos}, bucking the typical trend of \chandra\ measurements resulting in higher temperatures \citep{Wallbank, Fiona}. The resulting C-statistic is 1757 for 1689 degrees of freedom (dof).

For A2744, the 2T model returns $T_{\rm H} = 17.2^{+1.3}_{-1.5}$~keV and $T_{\rm L} = 2.9^{+0.4}_{-0.5}$~keV, giving a C-statistic of 1733 for 1647 dof. This represents a modest improvement relative to the 1T model and reinforces the view that the cluster contains strongly non-isothermal gas as observed in previous studies \citep[e.g.,][]{A2744Owers}.

The T+IC fit returns $kT = 11.5 \pm 0.5$~keV, consistent with the 1T model, as either a hot thermal tail or a non-thermal power law can explain the high-energy continuum. The model yields C-stat = 1743 for 1648 dof, worse than the 2T result, suggesting that the emission is dominated by thermal processes. The IC component is not significantly detected, and we derive an upper limit on its flux, as we will discuss in Section~\ref{subsec:NTE}.

Although a three-component model (2T+IC) is, in principle, the most complete physical description, the parameters become highly degenerate in practice. Either $T_{\rm L}$ is driven to unphysical values or the IC normalization becomes suppressed, and the fit effectively collapses into either a 2T or T+IC solution with comparable IC flux uncertainties. Thus, the additional freedom provides no meaningful improvement to the modeling. For a more complete discussion on this, see \citet{rojas21}.

Table~\ref{table:fits} summarizes the statistical comparison of the spectral models. Based solely on the nominal C-stat values, the two-temperature model provides the best fit, consistent with previous findings that massive merging clusters exhibit substantial thermal structure \citep{665ShockDasadia, Helen2146}. However, these values do not yet incorporate the known systematic uncertainties associated with the \nustar\ background.

The parameter uncertainties reported in Table~\ref{table:fits} are given as statistical uncertainties followed by systematic uncertainties. The statistical uncertainties reflect the constraints from the spectral fits, while the systematic uncertainties quantify the sensitivity of the best-fit parameters to background variations, as described below. The C-stat ranges represent the distributions obtained from the 1000 background realizations, rather than confidence intervals on the fitted parameters.

Although the 2T model performs the best among the options tested, the C-statistic alone does not quantify whether its additional thermal component is warranted \citep{Ftest}. We therefore applied the {\tt ftest} command in {\tt XSPEC} to compare the nested 1T and 2T models. For A2744, the resulting probability is 0.067, implying a 93.3\% likelihood that the improvement of the 2T model over the simpler thermal model is statistically meaningful. While the ICM is undoubtedly multi-phase, the quality and spectral resolution of the \nustar\ data allow a two-temperature model to capture the dominant features of the emission.

\begin{deluxetable*}{lccc}
\tabletypesize{\scriptsize}
\tablewidth{0pt}
\tablecaption{Best-fit spectral parameters for the 1T, 2T, and T+IC models of A2744. Where two uncertainties are listed, the first is statistical and the second is systematic. Systematic uncertainties are estimated using the background Monte Carlo analysis described in Section~\ref{subsubsec:best}. C-stat ranges show the distributions from the 1000 background realizations.}
\tablehead{
\colhead{Parameter} &
\colhead{1T} &
\colhead{2T} &
\colhead{T+IC}
}
\startdata
$kT$ (keV) & $10.9 \pm 0.5,\ 0.4$ & \nodata & $11.5 \pm 0.5,\ 0.8$ \\
$T_{\rm H}$ (keV) & \nodata & $17.2^{+1.3,+0.8}_{-1.5,-0.4}$ & \nodata \\
$T_{\rm L}$ (keV) & \nodata & $2.9^{+0.4,+0.2}_{-0.5,-0.1}$ & \nodata \\
$\Gamma$ & \nodata & \nodata & $2.43$ (fixed) \\
Abundance (solar) & $0.14 \pm 0.03,\ 0.01$ & $0.23^{+0.08,+0.05}_{-0.05,-0.07}$ & $0.22 \pm 0.05,\ 0.03$ \\
Norm$_{\rm T}$\tablenotemark{a} & $11.1 \pm 0.2,\ 0.1$ & \nodata & $5.5 \pm 0.2,\ 0.2$ \\
Norm$_{\rm TH}$\tablenotemark{a} & \nodata & $2.9^{+0.4,+0.8}_{-0.7,-0.9}$ & \nodata \\
Norm$_{\rm TL}$\tablenotemark{a} & \nodata & $9.9^{+0.9,+0.5}_{-1.0,-0.9}$ & \nodata \\
Norm$_{\rm IC}$\tablenotemark{b} & \nodata & \nodata & $1.58^{+0.48,+0.25}_{-0.41,-0.59}$ \\
Redshift $z$ & $0.319 \pm 0.17$ & $0.322 \pm 0.15$ & $0.314 \pm 0.15$ \\
C-stat\tablenotemark{c} & $1757^{+135}_{-132}$ & $1733^{+120}_{-125}$ & $1743^{+122}_{-115}$ \\
Degrees of freedom & 1649 & 1647 & 1648 \\
\enddata
\tablenotetext{a}{Normalization of the APEC model, given by $(10^{-14}/[4 \pi (1+z)^{2} D^{2}_{A}]) \int n_{e} n_{H} dV$, where $z$ is the redshift, $D_{A}$ is the angular diameter distance, $n_{e}$ is the electron density, $n_{H}$ is the ionized hydrogen density, and $V$ is the volume of the cluster. The normalizations are reported in units of $10^{-3}$~cm$^{-5}$.}
\tablenotetext{b}{Power-law normalization; the corresponding 20--80~keV IC flux is in units of $10^{-12}$~erg~s$^{-1}$~cm$^{-2}$.}
\tablenotetext{c}{Distribution of C-stat values from the 1000 background realizations shown.}
\label{table:fits}
\end{deluxetable*}

\subsubsection{Impact of Systematic Uncertainties}
\label{subsubsec:best}

\begin{figure}
\centering
\includegraphics[scale=0.20]{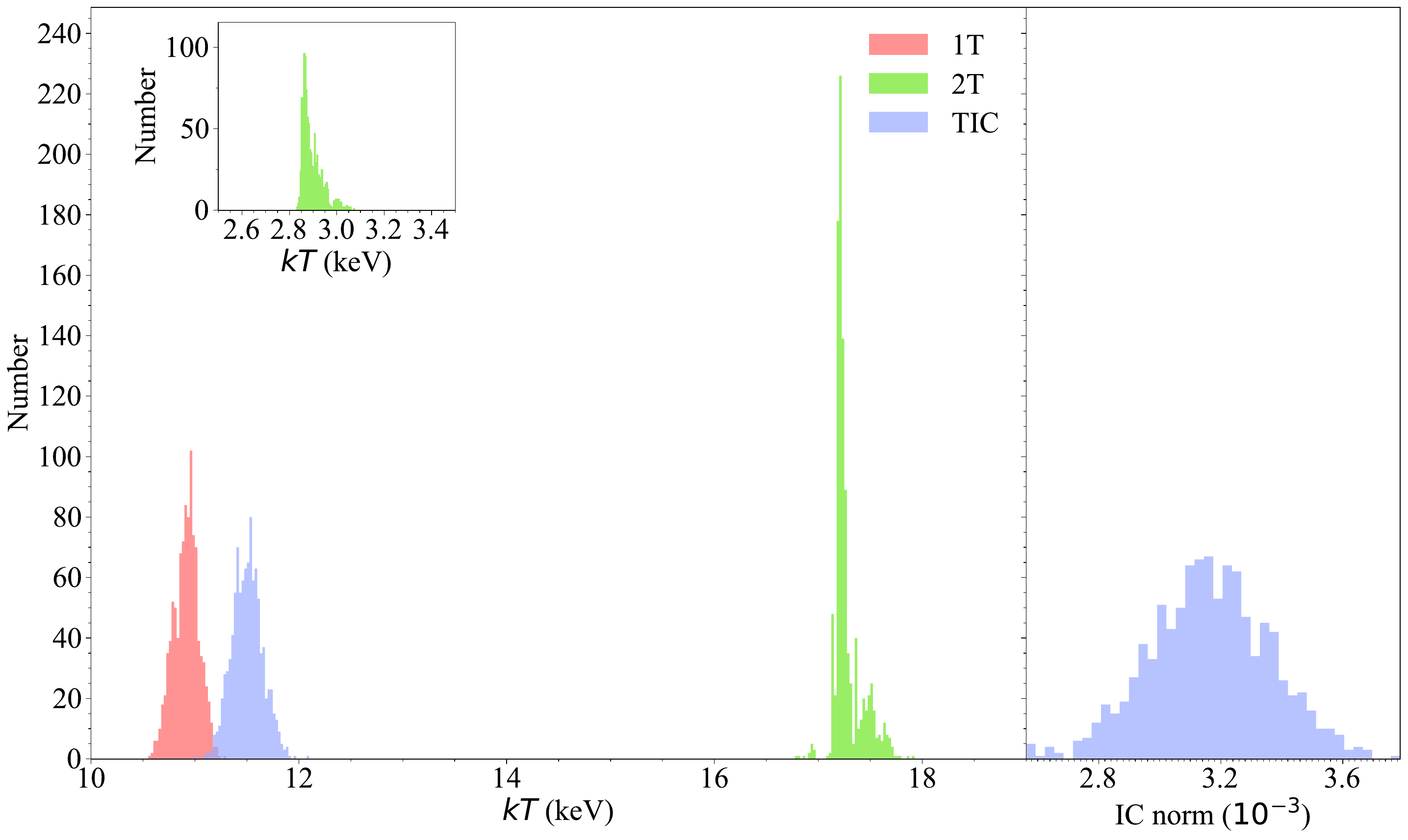}
\includegraphics[scale=0.60]{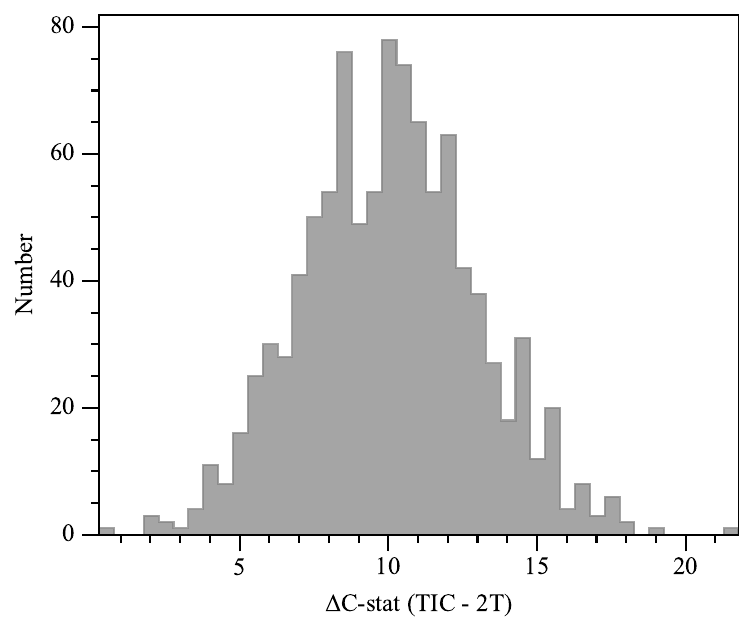}

\caption{Top: The distribution of best-fit parameter values for the 1T (red), 2T (green), and T$+$IC (blue) models using the 1000 realizations of the background (as described in Section~\ref{subsec:uncertainties}) for A2744. Parameters shown are the temperatures of each model and the IC norm. Bottom: The distribution of the difference in C-stat values ($\Delta$$C$) between the T$+$IC and 2T models from fits using the 1000 realizations of the background. The 2T model is preferred in all iterations for describing the \nustar-observed spectra over the T$+$IC model, with the latter not having any realization favoring it over the 2T model. Therefore, we conclude that the data clearly disfavor the addition of a non-thermal component.}
\label{fig:histograms}

\end{figure}

The \nustar\ background contains several components: instrumental continuum, aperture stray light, and focused cosmic background, each of which carries well-characterized normalization uncertainties (Appendix~\ref{subsec:uncertainties}). To assess their effect on the spectral modeling, we performed a Monte Carlo analysis in which the normalizations of the dominant components were perturbed using Gaussian deviates consistent with their systematic errors following the methodology in \citet{wikBullet} and \citet{rojas21}. For each of the 1000 realizations, we refit the spectra and recorded the resulting best-fit parameters. The parameter distributions for A2744 are shown in Figure~\ref{fig:histograms}.

The impact of background variations depends strongly on model complexity. In the 1T case, the continuum shape is governed almost entirely by the temperature, so perturbing the background produces only small deviations around the nominal best-fit value. This behavior appears in the narrow red distribution, indicating that systematic uncertainties remain comparable to the purely statistical ones.

The additional degrees of freedom in the 2T and T+IC models make them more responsive to background fluctuations. In the 2T model, the high-temperature component $T_{\rm H}$ is particularly sensitive: lowering the background shifts the spectral turnover to higher energies and drives $T_{\rm H}$ upward, whereas increasing the background has the opposite effect. The lower-temperature component $T_{\rm L}$ adjusts correspondingly to maintain agreement at the soft end of the bandpass, leading to the broader green distribution.

For the T+IC model, most of the response to background variation is absorbed by the power-law normalization. Because the shape of the IC component resembles that of the high-energy background, modest background shifts translate directly into changes in inferred IC flux. Nonetheless, even large perturbations produce only moderate variations in the IC normalization, and all realizations return fluxes consistent with a non-detection. In this sense, the IC component behaves much like $T_{\rm L}$ in the 2T fit, providing a flexible adjustment at the extremes of the energy range without representing a physically required component.

Allowing the background to vary within its systematic bounds does not alter the qualitative model preference. Across all 1000 realizations, the 2T model consistently achieves lower C-stat values than the T+IC model, with $\Delta$C-stat distributions centered near the nominal value of $\sim10$. No realization favors a statistically significant IC detection.

The residuals around $\sim20$~keV show excess scatter relative to the statistical errors, reflecting the fact that the Poisson uncertainties do not include background systematics. Imperfections in modeling the \nustar\ fluorescent lines contribute to these residuals. Incorporating known systematic uncertainties in quadrature would reconcile the apparent scatter with the expected deviations.

\section{Shock Front Analysis} \label{sec:tmap}

\subsection{Temperature Map}

To characterize the thermal structure of the southeastern sector of A2744 and determine the precise location of the potential shock front, we first constructed a temperature map of the cluster following the methodology in \citet{Cicely478} and \citet{ChristianA665}. Because shock identification relies on detecting localized temperature discontinuities, the cluster was partitioned into approximately isothermal regions through spatially resolved spectral fitting.

We generated the initial map by dividing the \nustar\ spectrum into three narrow energy ranges: $3-5$~keV, $6-10$~keV, and $10 - 20$~keV. Circular extraction regions were seeded on a grid at every fifth pixel, with an initial radius of 5 pixels. The radius of each region was then iteratively expanded until at least 1000 source counts were obtained, ensuring temperature measurements with better than 10\% statistical precision. This procedure was applied to the raw count, background, and exposure maps in each energy band.

For each circular region, the exposure-corrected and background-subtracted counts across the three bands were treated as a coarse spectrum and fit with a thermal model in {\tt XSPEC}, yielding a best-fit temperature. The resulting temperature value was assigned to the central pixel of the region, and a continuous map was produced by interpolating between these values across the field of view.

This procedure provides a qualitative view of the overall temperature distribution but has limitations in resolving sharp temperature discontinuities. The temperature estimates are derived from counts in only three energy bands, rather than detailed spectral fitting, and the interpolation between circular regions can smooth genuine temperature discontinuities or produce localized features that do not accurately reflect the underlying temperature structure. The limited signal-to-noise ratio, particularly at higher energies, can further affect the reliability of local temperature estimates.

The resulting \nustar\ temperature map of A2744 is shown in Figure 6. Although the map provides a general indication of the cluster's temperature structure, individual shock fronts are not clearly resolved. In particular, the apparent localized hot features in regions where cooler upstream gas would be expected may reflect limitations of the map-generation procedure and the available signal-to-noise ratio, and should not be interpreted as resolved shock structures. We therefore adopt the upstream and downstream regions defined by \citet{A2744Owers} for our quantitative shock analysis rather than identifying shock boundaries from the preliminary temperature map.

Our analysis focuses on the southeastern sector of A2744, where the available \nustar\ observations allow us to characterize the temperature discontinuity across the shock. Although a shock is also present in the northeastern outskirts, the limited signal-to-noise ratio in that region prevents us from obtaining temperature measurements with sufficient precision to constrain the corresponding temperature jump. We therefore restrict our quantitative shock analysis to the southeastern sector, where the data provide a more reliable basis for measuring the shock properties.

\subsection{Region Selection}\label{subsec:regionselection}

The extraction regions were defined by jointly examining the preliminary \nustar\ temperature map and the \chandra\ surface-brightness morphology from \citet{A2744Akos}. The temperature map provided a qualitative guide to the overall thermal structure and helped identify regions of approximately uniform temperature, while the \chandra\ image provided additional information on the spatial structure of the intracluster medium. However, because individual shock fronts were not clearly resolved in the preliminary temperature map, their locations were not determined from this map alone.

To characterize the temperature discontinuity across the southeastern shock front, we divided this sector into two extraction regions corresponding to the pre-shock (upstream) and post-shock (downstream) gas. The boundary between these regions was selected with reference to the X-ray surface-brightness edge associated with the shock identified by \citet{A2744Owers}. This shock aligns with the southeastern boundary of the radio halo shown in the radio images of \citet{A2744VLA} and is distinct from the separate, weaker shock associated with the southeastern radio relic.

While the upstream and downstream regions were defined to isolate the southeastern shock, other extraction regions across the cluster -- particularly those with significantly different temperatures -- remain within the \nustar\ point spread function (PSF) response. As a result, scattered photons from outside the shock regions may contaminate the measured spectra, motivating a detailed assessment of region-to-region crosstalk.

Region sizes and shapes were adjusted with this crosstalk in mind to ensure sufficient S/N for reliable spectral fitting with {\tt nucrossarf} (see the following subsection). Each extraction region was required to be at least $30\arcsec$ across, preventing individual regions from being dominated by PSF distributed emission and ensuring that cross-contamination between neighboring regions remained manageable. These regions and their corresponding temperatures are shown in Figure~\ref{fig:regionmap} and Table~\ref{table:t2}.

\subsection{Region Crosstalk}

\nustar\ provides the unique capability to focus hard X-rays above 10~keV, but this sensitivity comes with the trade-off of a relatively broad, weakly energy-dependent point spread function. Approximately half of all detected photons are distributed within a radius of $\sim$1' (the Half Power Diameter, HPD), while the core of the PSF has a Full Width at Half Maximum (FWHM) of $\sim$18". In extended, structured systems such as galaxy clusters, this broad PSF redistributes photons over large spatial scales, meaning that emission measured in any given region may include non-negligible contributions from elsewhere in the field of view.

This effect is especially important for shock-front diagnostics, where pre- and post-shock temperatures may differ significantly. Even a small fractional spill -- hot gas leaking into the upstream region or vice-versa -- can bias the measured temperature jump. Because standard ARF generation does not account for flux originating outside the extraction region (Section~\ref{sec:observe}), we calculated cross-contamination auxiliary response functions (cross-ARFs) using the {\tt nucrossarf}\footnote{\url{https://github.com/danielrwik/nucrossarf}}
 software \citep{Aysegul}. These routines compute the effective contribution of every region to every other region, producing a full cross-response matrix that captures PSF-mediated mixing across the cluster.
    
This analysis package treats source and spectral extraction regions independently, creating ARFs for a source at some location appropriate for all spectra given the regions position relative to the source and the energy-dependence of the PSF.
    
If the user creates $N$ spectral extraction regions around $N$ sources, the typical scenario for diffuse source analysis, $N \times N$ cross-ARFs will be created to take into account the wings of the PSF of each source within the other regions. 

Using the regions discussed in Section~\ref{subsec:regionselection}, cross-ARFs are generated to account for the scattered photons throughout the cluster and used to fit the spectra for each region to retrieve a more true temperature. The abundance and redshift obtained from our global fit to the cluster are used in all region fitting. The resulting fit parameters are shown in Table~\ref{table:t2}.

\begin{deluxetable}{ccc}
\tablewidth{0pt}
\tablecaption{Results from the nucrossarf Analysis (Section 4.3). 
}
\tablehead{
& \colhead{kT} & \colhead{Norm\tablenotemark{1}} \\
\colhead{Region} & \colhead{(keV)} & \colhead{($10^{-3}$)}
}
\startdata
1 & $9.52^{+1.5}_{-1.2}$  & $1.62 \pm 0.08$ \\
2 & $7.40^{+1.1}_{-0.9}$ & $4.08 \pm 0.05$ \\
3 & $6.62^{+0.7}_{-1.1}$ & $4.41 \pm 0.03$ \\
4 & $8.72^{+1.1}_{-1.2}$ & $1.56 \pm 0.03$ \\
5 & $7.93^{+1.6}_{-1.2}$ & $3.11 \pm 0.05$ \\
6 & $6.46^{+1.2}_{-1.4}$ & $2.28 \pm 0.07$ \\
7 & $10.20^{+1.5}_{-1.1}$ & $1.07 \pm 0.06$ \\
8 & $9.31^{+1.3}_{-1.1}$ & $2.71 \pm 0.03$ \\
9 & $8.16^{+0.9}_{-1.0}$ & $7.44 \pm 0.05$ \\
10 & $10.95^{+1.2}_{-1.7}$ & $1.19 \pm 0.04$ \\
11 & $12.15^{+1.7}_{-1.3}$ & $8.78 \pm 0.09$ \\
12 & $7.90^{+1.3}_{-1.2}$ & $6.59 \pm 0.08$ 
\enddata
\tablenotetext{1}{Normalization of the APEC model, given by $(10^{-14}/[4 \pi (1+z)^{2} D^{2}_{A}]) \int n_{e} n_{H} dV$ where $z$ is the redshift, $D_{A}$ is the angular diameter distance, $n_{e}$ is the electron density, $n_{H}$ is the ionized hydrogen density, and $V$ is the volume of the cluster.}
\label{table:t2}
\end{deluxetable}

\begin{figure*}
\centering
\includegraphics[scale=0.45]{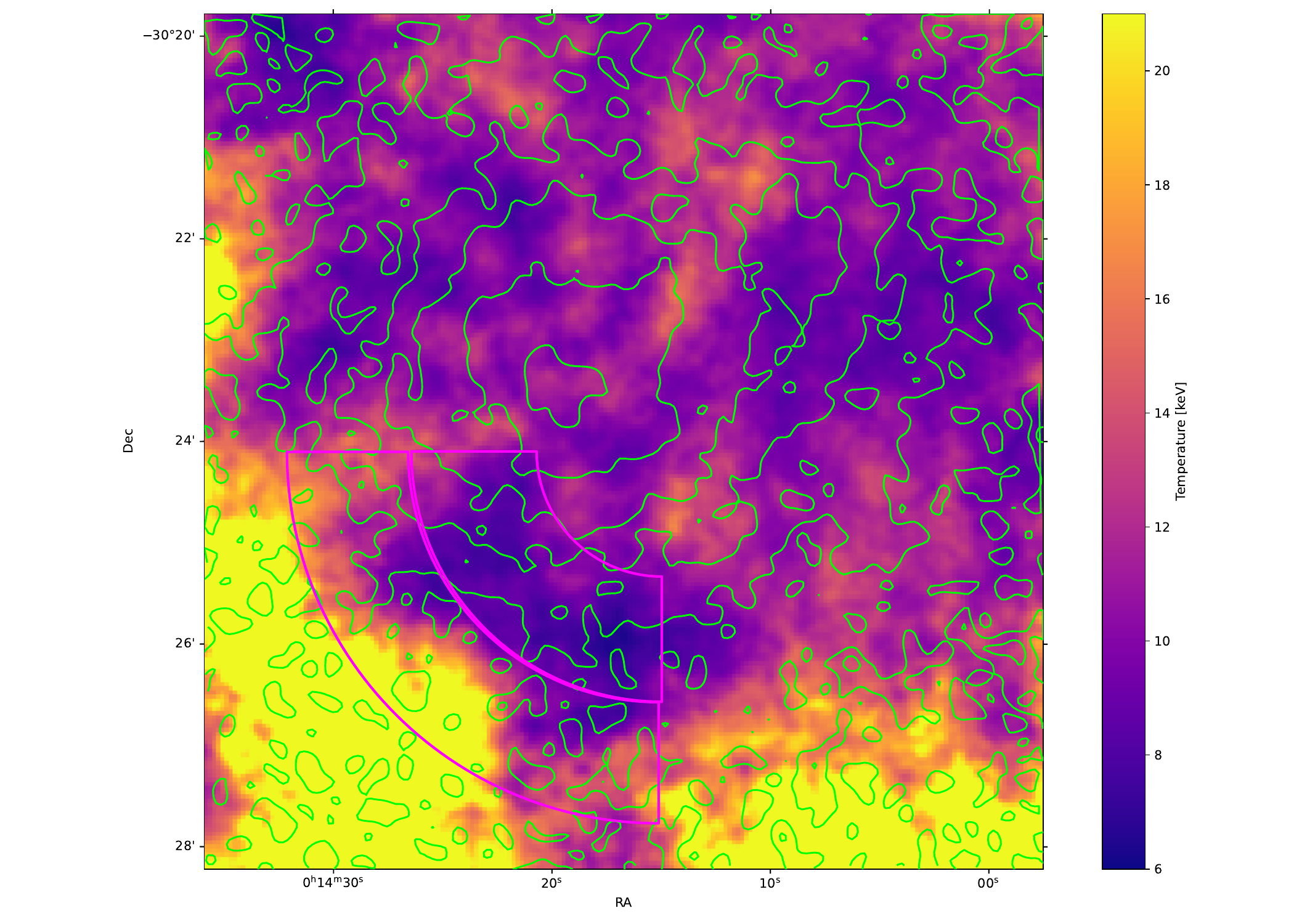}
\caption{Preliminary NuSTAR temperature map of A2744 used to inform the selection of extraction regions for the shock analysis. The upstream and downstream regions adopted from \citet{A2744Owers} are shown in magenta. The contours show the NuSTAR 3--20 keV image and were generated using DS9.}
\label{fig:tempmap}

\end{figure*}

\begin{figure*}
\centering

\includegraphics[scale=0.5]{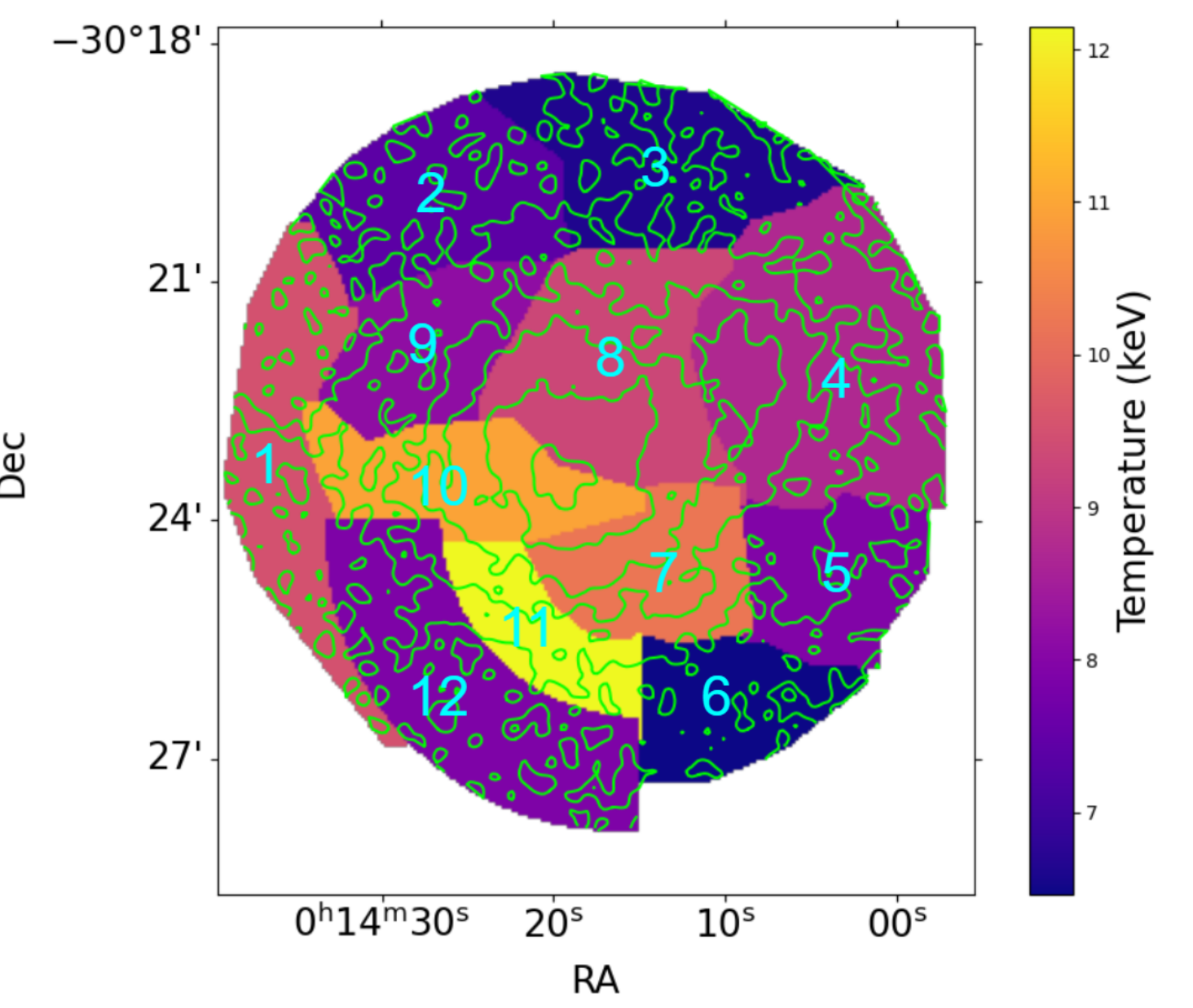}
\caption{Temperature map displaying results from the nuCrossARF analysis, corresponding to the
temperatures in Table 2. The color bar indicates the temperature in keV. X-ray contours are shown for reference.}
\label{fig:regionmap}

\end{figure*}

\section{Discussion} \label{sec:Summary}

\subsection{Non-thermal Emission} \label{subsec:NTE}

From our T$+$IC model we can set the 90\% upper limits $20-80$~keV flux of non-thermal emission coming from A2744 of $F_{\rm NT} < 2.31 \times 10^{-12}$~\flux. Although the NuSTAR spectra were fitted over 3--20~keV, this flux is calculated by integrating the best-fit power-law model over the 20--80~keV band. This energy range facilitates comparison with previous NuSTAR studies of inverse Compton emission in galaxy clusters \citep[e.g.,][]{wikBullet, rojas21, Rojas665}. Because the IC component is not significantly detected, this value represents an upper limit rather than a measured flux. 

Based on the statistical tests and comparisons described in Section~\ref{subsec:best}, we can safely rule out the presence of large amounts of IC scattering when comparing the 2T and T+IC models. The data is best fit by a two temperature component, purely thermal model. The variation in power law normalizations shown in the right panel of Figure~\ref{fig:histograms}, which translate to variations in IC flux, is due to the IC flux acting as an extra, lower temperature component in the T+IC model. The addition of this power law component does not fit the data as well as the 2T model as shown in C-stat comparisons the bottom panels of the previously mentioned figures. There are no previous IC flux measurements that have been published for A2744.\\

\subsection{Cluster Magnetic Field} \label{subsec:magnets}

As described in Section~\ref{sec:intro}, with an upper limit on the IC flux we can set a lower limit on the average magnetic field strength $B$ in the ICM. To do this, we must relate the synchrotron‐emitting relativistic electron population to its expected high-energy inverse Compton (IC) output. Turbulence generated during a major merger can stochastically (re)accelerate electrons, producing Mpc-scale diffuse radio halos \citep{reacceleration}. These same electrons, interacting with the CMB, should also upscatter CMB photons into the hard X-ray band through IC scattering.

Detecting this IC component -- or constraining it, as in the current analysis -- is critical for understanding the non-thermal energy budget of the cluster. If relativistic particles contribute non-negligible pressure support, hydrostatic mass estimates may be biased low \citep[e.g.,][]{1993ApJ...407L..49B, 2009astro2010S.305V, 2019A&A...621A..39E}. The ratio of the synchrotron radio flux ($F_R$) to the X-ray IC flux ($F_X$) directly encodes the balance of energy densities between the magnetic field and the CMB. For a single electron emitting both synchrotron and IC radiation, the flux ratio simplifies to the ratio of energy densities:

\begin{equation}
\label{eq:1}
\frac{F_R}{F_X} =
\frac{U_B}{U_{\rm CMB}} =
\frac{B^2/8\pi}{aT^4_{\rm CMB}} , .
\end{equation}

By generalizing this expression to a power-law distribution of electron energies, we can derive the magnetic field strength required to reproduce both the observed radio emission and the (non-)detection of IC emission. Following \cite{1979rpa..book.....R,1994hea..book.....L}, the resulting form is:

\begin{equation}
\label{eq:2}
B = C(p)(1+z)^{(p+5)/(p+1)} \times
\bigg(\frac{F_R}{F_X}\bigg)^{2/(p+1)}
\bigg(\frac{\nu_R}{\nu_X}\bigg)^{(p-1)/(p+1)} ,
\end{equation}

where $p$ is the electron energy index ($N(E) \propto E^{-p}$, with $p = 2\alpha + 1$ for a synchrotron spectrum of index $\alpha$), and $C(p)$ is a constant of proportionality. We adopt a spectral index of $\alpha = 1.43$ following \citet{A2744VLA}, giving $C(p) \approx 369$. We take $\nu_X = 2.418 \times 10^{17}$~Hz, corresponding to 1~keV, and integrate the IC model over the 20--80~keV band to obtain the X-ray flux. Since we measure only an upper limit on $F_X$, Equation~\ref{eq:2} provides a lower limit on the average magnetic field in the emitting volume.

Using Equation~\ref{eq:2} with our IC upper limit and the total diffuse radio flux of $49.7$~mJy measured at $1.4$~GHz within a global extraction region approximately matching that used by \citet{A2744VLA} for their radio flux measurement, we obtain a lower limit of $B \gtrsim 0.2~\mu$G for Abell~2744. For comparison, \citet{MagA2744} estimated $B \sim 0.5~\mu$G under the assumption of local equipartition between the relativistic particle energy density and the magnetic field energy density. Our constraint is therefore broadly consistent with previous estimates for other merging clusters \citep{wikBullet, rojas21, Rojas665, Abell523, Creech}. As is typical for IC-based limits, the inferred field strength remains below equipartition values but does not preclude them.

The significance of \nustar's ability to place these constraints is that robust IC upper limits directly challenge past claims of IC detections \citep{MillionIC}, which, if real, would imply magnetic fields well below equipartition. By ruling out bright IC emission, we maintain the possibility of magnetic fields stronger than our lower limit of $B \gtrsim 0.2~\mu$G. Typical cluster-wide magnetic field strengths are of order $1~\mu$G, while fields of tens of $\mu$G have been proposed as necessary for substantial magnetic pressure support of the intracluster medium \citep{CarilliTaylor2002}. Our constraint does not exclude such field strengths, although it does not establish that magnetic pressure contributes significantly to hydrostatic mass biases. In extreme cases, non-thermal pressure support can reach $\sim$20--30\% \citep{A3667,A3667NW}, comparable to the level needed to help reconcile the long-standing $\sim$20\% tension between hydrostatic and weak-lensing mass estimates \citep{Biffi}.

Rotation measure (RM) synthesis offers an independent route to magnetic field estimates and can yield field strengths of a few $\mu$G in cluster interiors \citep[e.g.,][]{Stuardi2021,Osinga2025}. For example, \citet{Stuardi2021} estimated a central magnetic field of $2.8\pm0.1~\mu$G within 200~kpc of Abell~2345, decreasing to $\sim0.3~\mu$G at the eastern radio relic. Similarly, \citet{Osinga2025} found an average field strength of $\sim2~\mu$G within $R_{500}$ across a sample of 124 clusters, assuming a magnetic-field fluctuation scale of 10~kpc. However, RM measurements are weighted by electron density and depend on the assumed magnetic-field structure, so these estimates are not directly equivalent to our IC-based lower limit. IC limits such as those presented here therefore provide a complementary constraint on cluster magnetic fields, particularly in lower-density regions where RM sensitivity is reduced and equipartition assumptions may break down.

\subsection{Shock Characterization}

To probe the dynamical impact of the merger in A2744, we focus on characterizing the southeastern shock originally reported by \citet{A2744Owers}, which aligns with the southeastern boundary of the radio halo \citep{A2744VLA}. This shock is distinct from the weaker shock associated with the southeastern radio relic, located approximately 0.5~Mpc away \citep{A2744VLA}. Characterizing the Mach number of the halo-edge shock provides an independent constraint on the thermodynamic impact of the merger and helps clarify the relationship between the X-ray shock structure and the cluster's diffuse radio emission. The Rankine-Hugoniot jump conditions relate the thermodynamic properties of gas across a shock front. For an ideal monatomic gas with adiabatic index $\gamma = 5/3$, the density ($\rho$) and temperature ($T$) jumps can be used to infer the Mach number (\mach) of the shock. Specifically, the Mach number can be expressed as a function of the density jump:

\begin{equation}
\mathcal{M} = \sqrt{\frac{4 (\rho_2 / \rho_1)}{(\gamma + 1) - (\gamma - 1)(\rho_2 / \rho_1)}},
\label{equation:den}
\end{equation}

or equivalently in terms of the temperature jump:

\begin{equation}
\mathcal{M} = \sqrt{\frac{(\gamma + 1)^2 \left(T_2/T_1 - 1 \right)}{2 \gamma (\gamma - 1)} + 1},
\label{equation:temp}
\end{equation}

where the subscripts 1 and 2 denote the pre-shock (upstream) and post-shock (downstream) regions, respectively.

The shock-front boundary is identified by a discontinuity in the density profile of the cluster. The pre- and post-shock densities are measured at this discontinuity and define the density jump. To derive the density profile, the cluster's X-ray surface brightness is fit with a parametric model assuming spherical symmetry. However, this approach is sensitive to the merger orientation: if the merger axis is tilted away from the plane of the sky, the apparent shock strength decreases \citep{Chadayammuri21}.

As an illustrative example, \citet{665ShockDasadia} measured a density jump in A665 of $\frac{\rho_{\rm post}}{\rho_{\rm pre}} = 3.2 \pm 0.3$ using Equation~\ref{equation:den}, corresponding to \mach\ = $3.4\pm0.7$. Modifying assumptions about the shock curvature reduced the density jump to $\frac{\rho_{\rm post}}{\rho_{\rm pre}} = 2.6$ (no errors reported), resulting in \mach\ = 2.5. These results demonstrate how geometric and modeling assumptions can influence the derived Mach number from density measurements.

Alternatively, the Mach number can be derived from the temperature jump across the shock. Spectral fitting of well-defined pre- and post-shock regions provides a direct measure of the thermal state of the gas, making this method less sensitive to projection or geometric uncertainties. While typically less precise than density-based estimates, temperature-derived Mach numbers are often more robust indicators of the true shock strength.

In the southeastern shock front, we measure a pre-shock temperature of $T_1 = 7.90^{+1.30}_{-1.20}$ keV and a post-shock temperature of $T_2 = 12.15^{+1.70}_{-1.30}$ keV (regions 12 and 11 in our temperature map respectively), enabling a direct application of Equation~\ref{equation:temp}. Using this temperature jump, we calculate a Mach number of \mach = $1.58_{-0.30}^{+0.42}$. For context, previous analyses of this shock using \chandra\ data measured \mach = $1.41_{-0.08}^{+0.11}$ using the density jump and \mach = $1.81_{-0.37}^{+0.49}$ using the temperature jump \citep{A2744Owers}, consistent within the errors to our \nustar\ result. The much deeper observations used in \citet{A2744Akos} found \mach = $1.44~\pm~0.01$ using the surface brightness jump and \mach = $1.69_{-0.37}^{+0.40}$ using the temperature jump, still showing consistent with our \nustar\ results.

Previous analyses of the southeastern region of A2744 reveal evidence for more than one shock front. In particular, \citet{A2744VLA} identified a weaker shock associated with the southeastern radio relic, located approximately 0.5~Mpc from the southeastern shock analyzed here. This secondary shock has a density-jump Mach number of $\mathcal{M}=1.26_{-0.15}^{+0.25}$ and lies outside the region where our NuSTAR temperature map provides reliable measurements. The deeper observations of \citet{A2744Akos} also identify this weaker shock, measuring Mach numbers of $\mathcal{M}=1.44\pm0.01$ and $\mathcal{M}=1.46_{-0.22}^{+0.25}$ from the surface brightness and temperature jumps, respectively. Our analysis therefore focuses on the southeastern shock reported by \citet{A2744Owers}, which aligns with the southeastern boundary of the radio halo, rather than the separate shock associated with the radio relic. The former may be associated with a merger event involving the main core and smaller substructures after core passage, whereas the latter likely formed during an earlier merger event.

\section{Summary}

\nustar\ observed A2744 during two observations with raw exposures of 130 ks and 59 ks, which after flare screening and manual filtering were reduced to 127 ks and 54 ks of clean time. Following background modeling and subtraction, we fit the cluster emission using three spectral models: a single-temperature thermal plasma model (1T), a two-temperature thermal plasma model (2T), and a single-temperature thermal model with an additional power-law component representing inverse Compton (IC) emission (T+IC). Allowing for systematic uncertainties in the background, the 2T model provides the best statistical fit to the broadband spectrum, capturing thermal structure that is not adequately described by a single plasma component. Models including a non-thermal IC component yield no improvement and are therefore disfavored. We conclude that the emission from A2744 is dominated by multi-temperature thermal gas, and we rule out the presence of a large IC flux within the region probed by \nustar.

Using the upper limit on the IC component from the T+IC fit, along with a total diffuse radio flux of 49.7~mJy at 1.4~GHz from \VLA\ measurements, we calculate a lower limit on the volume-averaged magnetic field of $B \gtrsim 0.2 \ \mu$G. This lies below but is consistent with the equipartition estimate of $B \sim 0.5 \ \mu$G reported previously for A2744 \citep{MagA2744}. While these values do not rule out the presence of stronger magnetic fields, they demonstrate that the relativistic electron population is insufficient to produce detectable IC emission at the sensitivity of our \nustar\ observation. For example, because the ratio of synchrotron to IC emission scales with the magnetic field energy density ($\propto B^{2}$), a fivefold increase in magnetic field strength would suppress the IC flux by a factor of $\sim25$ \citep{wikBullet}, requiring a correspondingly more sensitive hard X-ray telescope for detection. Even under these conditions, IC emission would only begin to compete with the thermal bremsstrahlung of a massive merging cluster in the $\sim30$--$50~\mathrm{keV}$ band, where the source flux is intrinsically faint and sensitivity is limited by background, implying that substantial gains would require increased effective area rather than deeper exposures alone. Despite these challenges, IC-based lower limits remain valuable because they constrain the magnetic field strength and help assess the potential dynamical importance of magnetic pressure in the intracluster medium. In localized regions where magnetic fields are amplified, magnetic pressure may approach the thermal pressure \citep{wikBullet}. Our results therefore demonstrate that the absence of a detectable IC component does not preclude dynamically significant magnetic fields in the intracluster medium.

To probe the merger dynamics in the southeastern sector, we characterize the shock front using the temperature discontinuity across the detected density edge. Spectral extraction across the boundary yields pre- and post-shock temperatures of $7.90^{+1.30}_{-1.20}$ and $12.15^{+1.70}_{-1.30}$~keV, which when inserted into the Rankine-Hugoniot relations give a shock strength of $\mathcal{M} = 1.58^{+0.42}_{-0.30}$. This is consistent with previous \chandra\ measurements (\mach $\sim$$1.41–1.81$ depending on method), confirming the presence of a moderately weak merger shock. A second southeastern shock, identified at a projected separation of $\sim$0.5 Mpc and exhibiting a lower Mach number, lies beyond the region where our \nustar\ temperature mapping remains reliable due to reduced S/N. Thus, our analysis focuses on the inner front, likely corresponding to the shock associated with the more recent core passage event.

\acknowledgments
\'A. B. and G.S. acknowledges support from the Smithsonian Institution and the Chandra Project through NASA contract NAS8-03060. This work made use of data from the \nustar\ mission, a project led by the California Institute of Technology, managed by the Jet Propulsion Laboratory, and funded by NASA.
We acknowledge support from NASA grant 80NSSC25K7084. This research has made use of the \nustar\ Data Analysis Software ({\tt NuSTARDAS}) jointly developed by the ASI Science Data Center (ASDC, Italy) and the California Institute of Technology (USA).

\typeout{}
\bibliography{mybibliography}

\appendix 

\section{Background Characterization and Modeling} \label{app:BGD}

\nustar\ is distinct among current X-ray observatories due to its open mast design, which physically separates the optics from the focal plane. Although this configuration enables hard X-ray focusing, it also allows unfocused photons to enter the detector plane directly and produce a stray-light gradient across the field of view. The background associated with this gradient must be modelled and separated from the more uniform instrumental background arising from particle and detector-related processes. In this Appendix, we describe how we construct our background models using the methods applied in \citet{wikBullet} and \citet{rojas21}, supported by {\tt nuskybgd}, as well as the physical origin of each background term.

\begin{figure}
    \centering
    \includegraphics[scale=0.40]{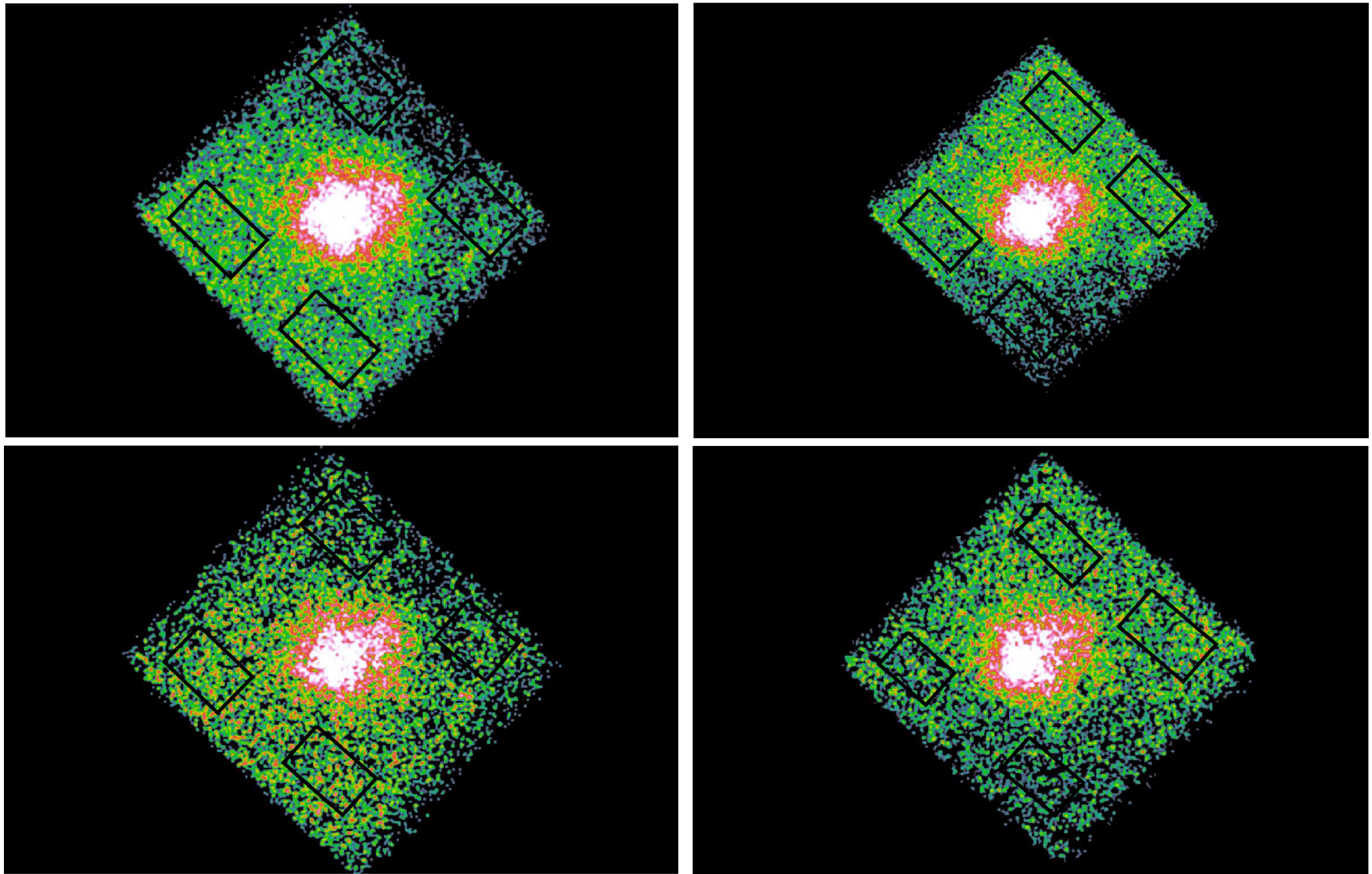}
    \caption{False color images of FPMA (left panels) and FPMB (right panels) for both observations of A2744 (OBSID 71001001002 on top, OBSID 71001001004 on bottom). Overlaid are the {\tt nuskybgd} regions used to extract background spectra from each of the four detectors. Background fits can be seen in Appendix~\ref{app:BGD}.}
    \label{fig:A665bgd}
\end{figure}

\begin{figure*}
\includegraphics[width=180mm]{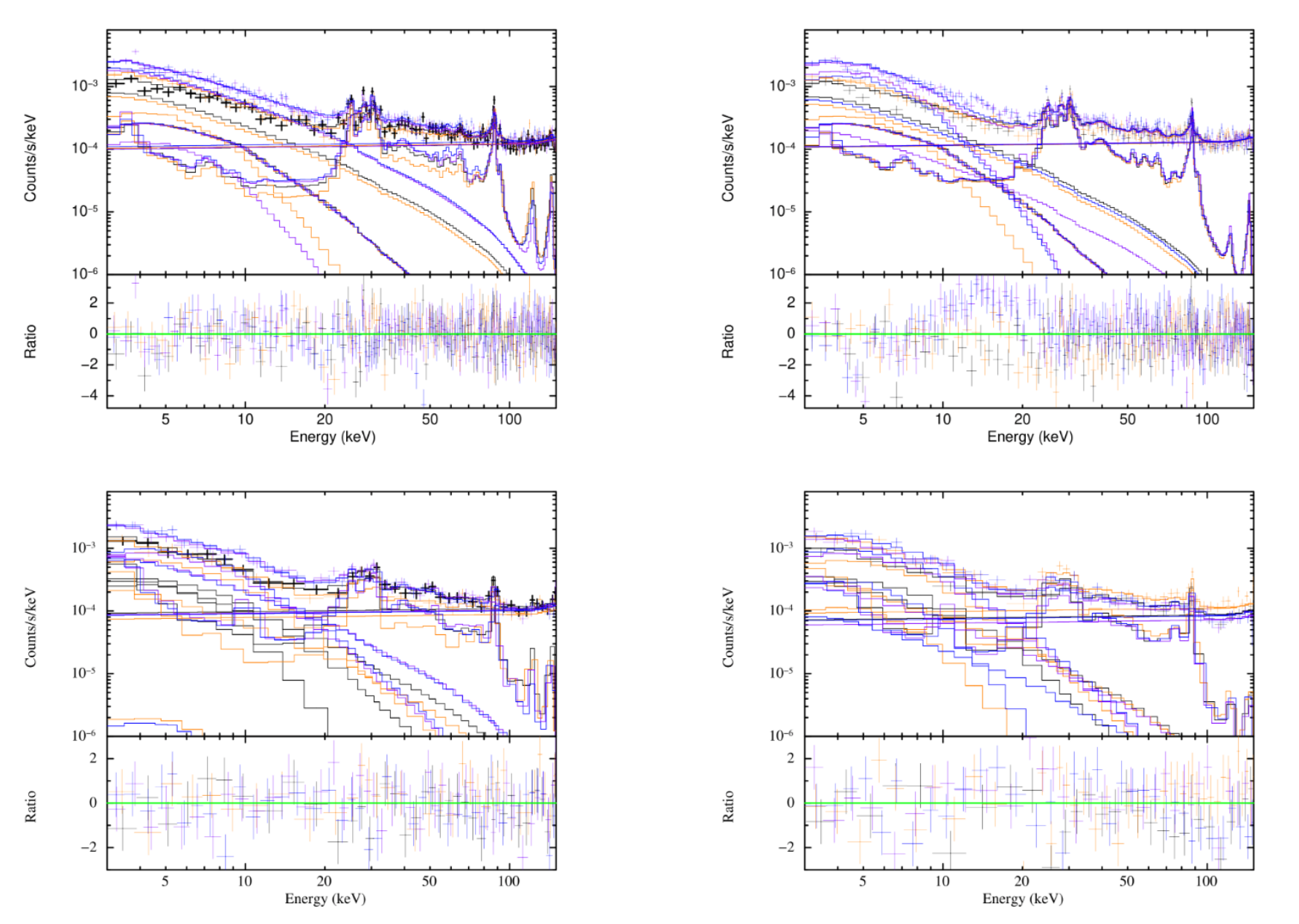}
\caption{
Fits to the background spectra extracted from the regions shown in Figure~\ref{fig:region}. The left and right panels show the spectra from FPMA and FPMB, respectively, while the top and bottom show the first observation (71001001002) and the second observation (71001001004) respectively. Each region is represented by a different color. Solar contributions produce the background at softer energies ($<$5~keV) which is combined with the model that includes instrumental background lines present at all energies. The undeflected CXB that passes through the aperture stops dominates between 10--20~keV, while the CXB that is focused through the optics (fCXB) is $\sim$10$\times$ fainter. ICM emission from the cluster produces the steeper spectral models in this energy range. At higher energies, cosmic ray-induced activation and fluorescence lines dominate, along with a flat continuum component representing the overall instrumental background level. For plotting purposes, adjacent bins are grouped until they have a significant detection at least as large as 10$\sigma$, with maximum 15 bins.}

\label{fig:bgdspec}

\end{figure*}

The \nustar\ background consists of multiple additive components that include internal detector background, aperture-stray light, reflected and scattered photons, and the focused cosmic X-ray background (fCXB). These terms are constrained using source-free detector regions, shown in Figure~\ref{fig:A665bgd}, which allow the relative contribution of each component to be determined and later extrapolated into cluster-emission regions. Because no region within the detector is truly free from cluster emission, we include an additional thermal contamination component described by an APEC model with a free temperature during background fitting. The resulting background spectral models for A2744 are shown in Figure~\ref{fig:bgdspec}. No significant differences were observed between FPMA and FPMB, indicating a consistent detector response across modules.

\subsection{Background Characterization} \label{subsec:components}
    
The \nustar\ background arises from several internal and external components that must be modeled carefully to extract faint cluster emission. A dominant contribution comes from the radiation environment encountered in low-Earth orbit, which drives the instrumental continuum level. In addition, activation and fluorescence lines — particularly strong between 22–32 keV — imprint narrow spectral features that must be accounted for when modeling the high-energy band. A further source is Compton scattering of gamma rays that penetrate the anti-coincidence shield and deposit energy in the detectors, producing additional internal background.

Because of its open mast structure, \nustar\ is also susceptible to stray light from off-axis cosmic X-ray background (CXB) photons. Ideally, the aperture stops on the focal plane bench would block these unfocused rays; however, mechanical limitations leave a $\sim$$3^\circ$  opening through which photons can enter directly. This produces a smooth spatial gradient across the detectors as the partially obstructing optics bench shadows the field of view. The precise pattern depends on telescope orientation and focal plane geometry. To account for this component, we adopt the {\tt HEAO-1 A2} CXB spectral model \citep{1987NASCP2464..339B}, valid over the 3–60 keV band.
   
    Another pitfall of the open design of \nustar~is that reflected and scattered X-rays from the entire sky can also strike the detectors. While many parts of the telescope can serve as a reflecting surface, the prime culprit is the backside of the aperture stops. Roughly 10-20$\%$ of the unfocused light that gets through is reflected \citep{wikBullet}. There are three main sources of reflected light: the Sun, the Earth, and the CXB. Terrestrial and solar light are primarily found around 1 keV during high periods of solar activity such as solar flares. 

    The focused cosmic background (fCXB) is the last component that we have to model. This is the contribution to the background from foreground and background sources that are unresolved within our field of view. These are mainly observed below 15 keV and correspond to roughly 10$\%$ of the photons from the CXB, with most of them entering unreflected through the aperture stops \citep{wikBullet}.

\subsection{Systematic Uncertainties} \label{subsec:uncertainties}

To evaluate potential inverse Compton emission, it is necessary to account for systematic uncertainty in each background component. Following \citet{wikBullet} and \citet{rojas21}, we adopt a 3\% uncertainty on the instrumental background, an 8\% uncertainty for the aperture CXB arising from cosmic variance across the sampled solid angle, and a much larger (up to 42\%) variability in the focused CXB component, reflecting field-to-field fluctuations in unresolved background source populations. These uncertainties are incorporated directly into our spectral fits so that faint non-thermal signatures cannot be misidentified due to background fluctuations. This approach ensures that upper limits derived on the IC flux remain conservative and robust.

\end{document}